\documentclass[12pt,a4paper]{article}
\usepackage{amssymb}
\usepackage{graphicx,color}
\makeatletter
\def\eqnarray{%
\stepcounter{equation}%
\let\@currentlabel=\theequation
\global\@eqnswtrue
\global\@eqcnt\z@
\tabskip\@centering
\let\\=\@eqncr
$$\halign to \displaywidth\bgroup\@eqnsel\hskip\@centering
$\displaystyle\tabskip\z@{##}$&\global\@eqcnt\@ne
\hfil$\displaystyle{{}##{}}$\hfil
&\global\@eqcnt\tw@$\displaystyle\tabskip\z@{##}$\hfil
\tabskip\@centering&\llap{##}\tabskip\z@\cr}
\makeatother
\newcommand{\ket}[1]{{\vert{#1}\rangle}}

\newcommand{\fukuso}{{\mathbf C}}

\newcommand{\seisu}{{\bf Z}}

\begin{document}

\title{\sl A New Algebraic Structure of Finite Quantum Systems and 
the Modified Bessel Functions}
\author{
  Kazuyuki FUJII
  \thanks{E-mail address : fujii@yokohama-cu.ac.jp }\\
  Department of Mathematical Sciences\\
  Yokohama City University\\
  Yokohama, 236-0027\\
  Japan
  }
\date{}
\maketitle
%\thispagestyle{empty}
%
%
%  gaiyou
%
%
\begin{abstract}
  In this paper we present a new algebraic structure (a super hyperbolic 
  system in our terminology) for finite quantum systems, which is 
  a generalization of the usual one in the two--level system. 
  
  It fits into the so--called generalized Pauli matrices, so they play 
  an important role in the theory. Some deep relation to the modified 
  Bessel functions of integer order is pointed out.
  
  By taking a skillful limit finite quantum systems become quantum mechanics 
  on the circle developed by Ohnuki and Kitakado.
\end{abstract}

\newpage

%
%
%     Honbun
%
%
Quantum Computation is usually based on two--level system of atoms (qubit 
theory). In the realistic construction of quantum logic gates we must 
solve some Schr{\" o}dinger equations. Then the Pauli matrices 
$\{\sigma_{1},\sigma_{3}\}$ is essentially used and not only the periodic 
functions $\{\cos(x),\sin(x)\}$ but also the hyperbolic functions 
$\{\cosh(x),\sinh(x)\}$ play an important role.

On the other hand, they are deeply related to the modified Bessel functions 
of integer order $\{I_{n}(x)\ |\ n\in \seisu\}$. The functions are in general 
given by the generating function.

Atom has usually many (finite or infinite) energy levels. However, to treat 
infinitely many ones at the same time is not realistic, so we treat an atom 
with finite (for example $n$) energy levels. We call this a finite quantum 
system and for this system the so--called generalized Pauli matrices 
$\{\Sigma_{1},\Sigma_{3}\}$ play a crucial role, see for example \cite{KF1}, 
\cite{KF2} and \cite{FFK}.

In this system we have a natural question on what functions corresponding to 
the hyperbolic functions are. In the paper we present such a system 
$\{c_{0}(x),c_{1}(x),\cdots,c_{n-1}(x)\}$ (a super hyperbolic system in our 
terminology) as a ``natural" generalization of $\{\cosh(x),\sinh(x)\}$.

Moreover, we define a generating matrix based on the generalized Pauli 
matrices as a ``natural" generalization of the generating function 
and obtain interesting results by taking some traces.

Lastly, we want to take a limit of finite quantum systems, which is of course 
impossible. However, there is a bypass. That is, by taking a skillful limit 
finite quantum systems become quantum mechanics on the circle developed by 
Ohnuki and Kitakado \cite{OK}.

Through this paper we have a clear and unified picture of quantum systems.

\par \vspace{5mm}
First of all we make some mathematical preliminaries on the $2$--level 
system. Let $\{\sigma_{1}, \sigma_{2}, \sigma_{3}\}$ be Pauli matrices and 
${\bf 1}_{2}$ the unit matrix : 
\begin{equation}
\sigma_{1} = 
\left(
  \begin{array}{cc}
    0 & 1 \\
    1 & 0
  \end{array}
\right), \quad 
\sigma_{2} = 
\left(
  \begin{array}{cc}
    0 & -i \\
    i & 0
  \end{array}
\right), \quad 
\sigma_{3} = 
\left(
  \begin{array}{cc}
    1 & 0 \\
    0 & -1
  \end{array}
\right), \quad 
{\bf 1}_{2} = 
\left(
  \begin{array}{cc}
    1 & 0 \\
    0 & 1
  \end{array}
\right).
\end{equation}

List the well--known properties of $\sigma_{1}$ and $\sigma_{3}$ : 
\begin{equation}
\sigma_{1}^{2}=\sigma_{3}^{2}={\bf 1}_{2},\quad 
\sigma_{1}^{\dagger}=\sigma_{1}, \quad 
\sigma_{3}^{\dagger}=\sigma_{3}, \quad 
\sigma_{3}\sigma_{1}=-\sigma_{1}\sigma_{3}=
\mbox{e}^{\pi i}\sigma_{1}\sigma_{3}.
\end{equation}

\par \noindent
Let $W$ be the Walsh--Hadamard matrix 
\begin{equation}
\label{eq:2-Walsh-Hadamard}
W=\frac{1}{\sqrt{2}}
\left(
  \begin{array}{cc}
    1 &  1 \\
    1 & -1
  \end{array}
\right)
=W^{-1}\ , 
\end{equation}
then we can diagonalize $\sigma_{1}$ as $\sigma_{1}=W\sigma_{3}W^{-1}$ 
by making use of $W$. 

The modified Bessel functions of integer order $\{I_{k}(x)\ |\ k\in \seisu\}$ 
are given by the generating function
\begin{equation}
\label{eq:generating function}
\mbox{e}^{\frac{x}{2}(w+\frac{1}{w})}=\sum_{k\in \seisu}I_{k}(x)w^{k}.
\end{equation}
Now let us list some (well--known) important properties (see for example 
\cite{WW}) :
\begin{eqnarray*}
1&=&I_{0}(x)+2\sum_{k=1}^{\infty}(-1)^{k}I_{2k}(x), \\
\mbox{e}^{x}&=&I_{0}(x)+2\sum_{k=1}^{\infty}I_{k}(x),\quad 
\mbox{e}^{-x}=I_{0}(x)+2\sum_{k=1}^{\infty}(-1)^{k}I_{k}(x) \\
\cosh(x)&=&I_{0}(x)+2\sum_{k=1}^{\infty}I_{2k}(x),\quad 
\sinh(x)=2\sum_{k=1}^{\infty}I_{2k-1}(x).
\end{eqnarray*}

In the following we set 
\begin{equation}
\label{eq:cosh-sinh}
c_{0}(x)\equiv \cosh(x)=\sum_{k=0}^{\infty}\frac{x^{2k}}{(2k)!},\quad 
c_{1}(x)\equiv \sinh(x)=\sum_{k=0}^{\infty}\frac{x^{2k+1}}{(2k+1)!}
\end{equation}
for simplicity. The fundamental equation 
\begin{equation}
\label{eq:fundamental identity two}
c_{0}^{2}(x)-c_{1}^{2}(x)=1
\end{equation}
is interpreted as a simple relation
\[
S\sigma_{3}S=\sigma_{3}\ \Longleftrightarrow\ 
\sigma_{3}S\sigma_{3}S={\bf 1}_{2}\ \Longleftrightarrow\ 
(\sigma_{3}S)^{2}={\bf 1}_{2}
\]
for $S$ defined by
\begin{equation}
S=
\left(
  \begin{array}{cc}
    c_{0}(x) & c_{1}(x) \\
    c_{1}(x) & c_{0}(x)
  \end{array}
\right)
=c_{0}(x){\bf 1}_{2}+c_{1}(x)\sigma_{1}=\mbox{e}^{x\sigma_{1}}.
\end{equation}

\vspace{10mm}
Next we would like to extend the $2$--level system to general $n$--level one. 
To make our purpose clearer we treat the $3$--level case in detail. 
Let $\sigma$ be $\mbox{exp}(\frac{2\pi i}{3})$, then we have
\begin{equation}
\label{eq:simple-relations}
\sigma^{3}=1,\quad \bar{\sigma}=\sigma^{2},\quad 1+\sigma+\sigma^{2}=0.
\end{equation}
Let $\Sigma_{1}$ and $\Sigma_{3}$ be generators 
of generalized Pauli matrices in the case of $n=3$, namely 
\begin{equation}
\Sigma_{1}=
\left(
  \begin{array}{ccc}
    0 &   & 1   \\
    1 & 0 &     \\
      & 1 & 0
  \end{array}
\right), \quad 
\Sigma_{3}=
\left(
  \begin{array}{ccc}
    1 &        &            \\
      & \sigma &            \\
      &        & \sigma^{2}
  \end{array}
\right).
\end{equation}
Then it is easy to see 
\begin{equation}
\Sigma_{1}^{3}=\Sigma_{3}^{3}={\bf 1}_{3},\quad 
\Sigma_{1}^{\dagger}=\Sigma_{1}^{2}, \quad 
\Sigma_{3}^{\dagger}=\Sigma_{3}^{2}, \quad 
\Sigma_{3}\Sigma_{1}=\sigma \Sigma_{1}\Sigma_{3}.
\end{equation}

Now we can show that $\Sigma_{1}$ can be diagonalized by making use of the 
matrix
\begin{equation}
\label{eq:3-Walsh-Hadamard}
W=\frac{1}{\sqrt{3}}
\left(
  \begin{array}{ccc}
    1 & 1          & 1          \\
    1 & \sigma^{2} & \sigma     \\
    1 & \sigma     & \sigma^{2}
  \end{array}
\right)\quad \in \quad U(3)
\end{equation}
like
\begin{equation}
\label{eq:3-diagonalization}
\Sigma_{1}=W\Sigma_{3}W^{\dagger}=W\Sigma_{3}W^{-1}.
\end{equation}
In fact 
\[
W\Sigma_{3}W^{\dagger}
=\frac{1}{3}
\left(
\begin{array}{ccc}
1 & 1          & 1          \\
1 & \sigma^{2} & \sigma     \\
1 & \sigma     & \sigma^{2}
\end{array}
\right)
\left(
\begin{array}{ccc}
1 &        &            \\
  & \sigma &            \\
  &        & {\sigma}^2 
\end{array}
\right)
\left(
\begin{array}{ccc}
1 & 1          & 1          \\
1 & \sigma     & \sigma^{2} \\
1 & \sigma^{2} & \sigma
\end{array}
\right)
=\frac{1}{3}
\left(
\begin{array}{ccc}
0 & 0 & 3  \\
3 & 0 & 0  \\
0 & 3 & 0
\end{array}
\right) 
=\Sigma_{1},
\]
where we have used the relations in (\ref{eq:simple-relations}). 

\vspace{3mm}
From (\ref{eq:cosh-sinh}) we set
\begin{equation}
c_{0}(x)=\sum_{k=0}^{\infty}\frac{x^{3k}}{(3k)!},\quad 
c_{1}(x)=\sum_{k=0}^{\infty}\frac{x^{3k+1}}{(3k+1)!},\quad 
c_{2}(x)=\sum_{k=0}^{\infty}\frac{x^{3k+2}}{(3k+2)!}.
\end{equation}
Then it is easy to check
\begin{equation}
\label{eq:defining formulas}
c_{0}(x)=
\frac{\mbox{e}^{x}+\mbox{e}^{\sigma x}+\mbox{e}^{\sigma^{2}x}}{3},\quad 
c_{1}(x)=
\frac{\mbox{e}^{x}+\sigma^{2}\mbox{e}^{\sigma x}+\sigma\mbox{e}^{\sigma^{2}x}}
{3},\quad 
c_{2}(x)=
\frac{\mbox{e}^{x}+\sigma\mbox{e}^{\sigma x}+\sigma^{2}\mbox{e}^{\sigma^{2}x}}
{3}
\end{equation}
by use of $\sigma$ in (\ref{eq:simple-relations}) or reversely
\[
\mbox{e}^{x}=c_{0}(x)+c_{1}(x)+c_{2}(x),\quad 
\mbox{e}^{\sigma x}=c_{0}(x)+\sigma c_{1}(x)+\sigma^{2}c_{2}(x),\quad 
\mbox{e}^{\sigma^{2}x}=c_{0}(x)+\sigma^{2}c_{1}(x)+\sigma c_{2}(x).
\]

Now, our question is as follows : What is the fundamental equation that 
$\{c_{0}(x),c_{1}(x),c_{2}(x)\}$ satisfy ?

\par \vspace{3mm} \noindent
The answer is given by the equation
\begin{eqnarray*}
&&(c_{0}(x)+c_{1}(x)+c_{2}(x))
(c_{0}(x)+\sigma c_{1}(x)+\sigma^{2}c_{2}(x))
(c_{0}(x)+\sigma^{2}c_{1}(x)+\sigma c_{2}(x)) \\
&&=\mbox{e}^{x}\mbox{e}^{\sigma x}\mbox{e}^{\sigma^{2}x}
=\mbox{e}^{(1+\sigma+\sigma^{2})x}
=\mbox{e}^{0}=1.
\end{eqnarray*}
By expanding the left-hand side and using the relations 
(\ref{eq:simple-relations}) we obtain
\begin{equation}
\label{eq:fundamental identity three}
c_{0}^{3}(x)+c_{1}^{3}(x)+c_{2}^{3}(x)-3c_{0}(x)c_{1}(x)c_{2}(x)=1.
\end{equation}

Next let us consider the addition formulas. By expanding
\[
\mbox{e}^{\sigma x}\mbox{e}^{\sigma y}=\mbox{e}^{\sigma (x+y)}
\quad \Longleftarrow \quad 
\mbox{e}^{\sigma t}=c_{0}(t)+\sigma c_{1}(t)+\sigma^{2}c_{2}(t)
\]
we have
\begin{eqnarray}
\label{eq:addition formulas}
&&c_{0}(x)c_{0}(y)+c_{1}(x)c_{2}(y)+c_{2}(x)c_{1}(y)=c_{0}(x+y), \nonumber \\
&&c_{0}(x)c_{1}(y)+c_{1}(x)c_{0}(y)+c_{2}(x)c_{2}(y)=c_{1}(x+y), \nonumber \\
&&c_{0}(x)c_{2}(y)+c_{1}(x)c_{1}(y)+c_{2}(x)c_{0}(y)=c_{2}(x+y).
\end{eqnarray}

From here let us give a unified approach by use of the generalized Pauli 
matrices $\{\Sigma_{1},\Sigma_{3}\}$ above. We consider the matrix
\begin{equation}
\mbox{e}^{x\Sigma_{1}}
=c_{0}(x){\bf 1}_{3}+c_{1}(x)\Sigma_{1}+c_{2}(x)\Sigma_{1}^{2}
=
\left(
\begin{array}{ccc}
c_{0}(x) & c_{2}(x) & c_{1}(x)  \\
c_{1}(x) & c_{0}(x) & c_{2}(x)  \\
c_{2}(x) & c_{1}(x) & c_{0}(x) 
\end{array}
\right).
\end{equation}
Then by $\Sigma_{1}=W\Sigma_{3}W^{\dagger}$ in (\ref{eq:3-diagonalization})
\[
\left(
\begin{array}{ccc}
c_{0}(x) & c_{2}(x) & c_{1}(x)  \\
c_{1}(x) & c_{0}(x) & c_{2}(x)  \\
c_{2}(x) & c_{1}(x) & c_{0}(x) 
\end{array}
\right)
=
\mbox{e}^{x\Sigma_{1}}=W\mbox{e}^{x\Sigma_{3}}W^{\dagger}
=
W
\left(
\begin{array}{ccc}
\mbox{e}^{x} &         &             \\
  & \mbox{e}^{x\sigma} &             \\
  &         & \mbox{e}^{x\sigma^{2}}
\end{array}
\right)
W^{\dagger},
\]
so taking the determinant leads to
\[
\left|
\begin{array}{ccc}
c_{0}(x) & c_{2}(x) & c_{1}(x)  \\
c_{1}(x) & c_{0}(x) & c_{2}(x)  \\
c_{2}(x) & c_{1}(x) & c_{0}(x) 
\end{array}
\right|
=
\mbox{e}^{(1+\sigma+\sigma^{2})x}=1.
\]
Namely, we recovered (\ref{eq:fundamental identity three}).

On the other hand, by use of (\ref{eq:3-Walsh-Hadamard}) it is 
straightforward to show
\begin{eqnarray*}
&&W
\left(
\begin{array}{ccc}
\mbox{e}^{x} &         &             \\
  & \mbox{e}^{x\sigma} &             \\
  &         & \mbox{e}^{x\sigma^{2}}
\end{array}
\right)
W^{\dagger}
=\frac{1}{3}
\left(
  \begin{array}{ccc}
    1 & 1 & 1                \\
    1 & \sigma^{2} & \sigma  \\
    1 & \sigma & \sigma^{2}
  \end{array}
\right)
\left(
\begin{array}{ccc}
\mbox{e}^{x} &         &             \\
  & \mbox{e}^{x\sigma} &             \\
  &         & \mbox{e}^{x\sigma^{2}}
\end{array}
\right)
\left(
  \begin{array}{ccc}
    1 & 1 & 1                \\
    1 & \sigma & \sigma^{2}  \\
    1 & \sigma^{2} & \sigma
  \end{array}
\right) \\
&&=
\frac{\mbox{e}^{x}+\mbox{e}^{\sigma x}+\mbox{e}^{\sigma^{2}x}}{3}{\bf 1}_{3}+
\frac{\mbox{e}^{x}+\sigma^{2}\mbox{e}^{\sigma x}+\sigma\mbox{e}^{\sigma^{2}x}}
{3}\Sigma_{1}+
\frac{\mbox{e}^{x}+\sigma\mbox{e}^{\sigma x}+\sigma^{2}\mbox{e}^{\sigma^{2}x}}
{3}\Sigma_{1}^{2},
\end{eqnarray*}
so we recovered (\ref{eq:defining formulas}).

The matrix form is very convenient. Moreover, we can give new relations. 
For that we consider the simple equation
\begin{equation}
\label{eq:fundamental additivity}
\mbox{e}^{x\Sigma_{1}}\mbox{e}^{y\Sigma_{1}^{\dagger}}
=\mbox{e}^{x\Sigma_{1}+y\Sigma_{1}^{\dagger}}.
\end{equation}
The left hand side is
\begin{eqnarray*}
\mbox{e}^{x\Sigma_{1}}\mbox{e}^{y\Sigma_{1}^{\dagger}}
&=&
(c_{0}(x){\bf 1}_{3}+c_{1}(x)\Sigma_{1}+c_{2}(x)\Sigma_{1}^{2})
(c_{0}(y){\bf 1}_{3}+c_{1}(y)\Sigma_{1}^{2}+c_{2}(y)\Sigma_{1}) \\
&=&
(c_{0}(x)c_{0}(y)+c_{1}(x)c_{1}(y)+c_{2}(x)c_{2}(y)){\bf 1}_{3} \\
&+&(c_{0}(x)c_{2}(y)+c_{1}(x)c_{0}(y)+c_{2}(x)c_{1}(y))\Sigma_{1} \\
&+&(c_{0}(x)c_{1}(y)+c_{1}(x)c_{2}(y)+c_{2}(x)c_{0}(y))\Sigma_{1}^{2}
\end{eqnarray*}
because $\Sigma_{1}^{\dagger}=\Sigma_{1}^{2}$. The right hand side is
\begin{eqnarray*}
\mbox{e}^{x\Sigma_{1}+y\Sigma_{1}^{\dagger}}
&=&
\mbox{e}^{W(x\Sigma_{3}+y\Sigma_{3}^{2})W^{\dagger}}
=
W\mbox{e}^{x\Sigma_{3}+y\Sigma_{3}^{2}}W^{\dagger}
=
W
\left(
\begin{array}{ccc}
\mbox{e}^{x+y} &         &                    \\
  & \mbox{e}^{x\sigma+y\sigma^{2}} &          \\
  &         & \mbox{e}^{x\sigma^{2}+y\sigma}
\end{array}
\right)
W^{\dagger}\\
&=&
\frac{\mbox{e}^{x+y}+\mbox{e}^{x\sigma+y\sigma^{2}}+
\mbox{e}^{x\sigma^{2}+y\sigma}}{3}{\bf 1}_{3}+
\frac{\mbox{e}^{x+y}+\sigma^{2}\mbox{e}^{x\sigma+y\sigma^{2}}+
\sigma\mbox{e}^{x\sigma^{2}+y\sigma}}{3}\Sigma_{1} \\
&+&
\frac{\mbox{e}^{x+y}+\sigma\mbox{e}^{x\sigma+y\sigma^{2}}+
\sigma^{2}\mbox{e}^{x\sigma^{2}+y\sigma}}{3}\Sigma_{1}^{2},
\end{eqnarray*}
so we obtain
\begin{eqnarray}
&&c_{0}(x)c_{0}(y)+c_{1}(x)c_{1}(y)+c_{2}(x)c_{2}(y)
=\frac{\mbox{e}^{x+y}+\mbox{e}^{x\sigma+y\sigma^{2}}+
\mbox{e}^{x\sigma^{2}+y\sigma}}{3}, \nonumber \\
&&c_{0}(x)c_{2}(y)+c_{1}(x)c_{0}(y)+c_{2}(x)c_{1}(y)
=\frac{\mbox{e}^{x+y}+\sigma^{2}\mbox{e}^{x\sigma+y\sigma^{2}}+
\sigma\mbox{e}^{x\sigma^{2}+y\sigma}}{3}, \nonumber \\
&&c_{0}(x)c_{1}(y)+c_{1}(x)c_{2}(y)+c_{2}(x)c_{0}(y)
=\frac{\mbox{e}^{x+y}+\sigma\mbox{e}^{x\sigma+y\sigma^{2}}+
\sigma^{2}\mbox{e}^{x\sigma^{2}+y\sigma}}{3}.
\end{eqnarray}

Next, let us consider the matrix 
$\mbox{e}^{x\Sigma_{1}+y\Sigma_{1}^{\dagger}}$. If we set $y=1/x$, then 
the matrix $\mbox{e}^{x\Sigma_{1}+(1/x)\Sigma_{1}^{\dagger}}$ is similar to 
(\ref{eq:generating function}) the generating function of modified Bessel 
functions of integer order. Therefore from (\ref{eq:generating function}) 
it is reasonable to consider
\begin{equation}
\mbox{e}^{\frac{x}{2}\left(w\Sigma_{1}+\frac{1}{w}\Sigma_{1}^{\dagger}\right)}
=
\mbox{e}^{\frac{x}{2}\left(w\Sigma_{1}+\frac{1}{w}\Sigma_{1}^{-1}\right)}
=
\sum_{k\in \seisu}I_{k}(x)w^{k}\Sigma_{1}^{k}.
\end{equation}
In the following we call this the {\bf generating matrix} of modified Bessel 
functions of integer order. Let us look for some typical properties. 
The result is
\begin{eqnarray}
\frac{1}{3}\mbox{tr}\left\{
\mbox{e}^{\frac{x}{2}\left(w\Sigma_{1}+\frac{1}{w}\Sigma_{1}^{\dagger}\right)}
\right\}
&=&
\frac{
\mbox{e}^{\frac{x}{2}(w+\frac{1}{w})}+
\mbox{e}^{\frac{x}{2}(w\sigma+\frac{1}{w}\sigma^{2})}+
\mbox{e}^{\frac{x}{2}(w\sigma^{2}+\frac{1}{w}\sigma)}
}{3}
=
\sum_{k\in \seisu}I_{3k}(x)w^{3k}, \nonumber \\
\frac{1}{3}\mbox{tr}\left\{
\mbox{e}^{\frac{x}{2}\left(w\Sigma_{1}+\frac{1}{w}\Sigma_{1}^{\dagger}\right)}
\Sigma_{1}\right\}
&=&
\frac{
\mbox{e}^{\frac{x}{2}(w+\frac{1}{w})}+
\sigma\mbox{e}^{\frac{x}{2}(w\sigma+\frac{1}{w}\sigma^{2})}+
\sigma^{2}\mbox{e}^{\frac{x}{2}(w\sigma^{2}+\frac{1}{w}\sigma)}
}{3}
=
\sum_{k\in \seisu}I_{3k-1}(x)w^{3k-1}, \nonumber \\
\frac{1}{3}\mbox{tr}\left\{
\mbox{e}^{\frac{x}{2}\left(w\Sigma_{1}+\frac{1}{w}\Sigma_{1}^{\dagger}\right)}
\Sigma_{1}^{2}\right\}
&=&
\frac{
\mbox{e}^{\frac{x}{2}(w+\frac{1}{w})}+
\sigma^{2}\mbox{e}^{\frac{x}{2}(w\sigma+\frac{1}{w}\sigma^{2})}+
\sigma\mbox{e}^{\frac{x}{2}(w\sigma^{2}+\frac{1}{w}\sigma)}
}{3}
=
\sum_{k\in \seisu}I_{3k-2}(x)w^{3k-2} \nonumber \\
&{}&
\end{eqnarray}
where $\sigma^{-1}=\sigma^{2}$ and $\sigma^{-2}=\sigma$.

A comment is in order. In the case of $n=2$ the generating matrix is
\[
\mbox{e}^{\frac{x}{2}\left(w\sigma_{1}+\frac{1}{w}\sigma_{1}^{\dagger}\right)}
=
\mbox{e}^{\frac{x}{2}\left(w+\frac{1}{w}\right)\sigma_{1}}
=
\cosh\left(\frac{x}{2}\left(w+\frac{1}{w}\right)\right){\bf 1}_{2}+
\sinh\left(\frac{x}{2}\left(w+\frac{1}{w}\right)\right)\sigma_{1}
\]
because $\sigma_{1}$ is hermitian, so the situation becomes much easier.

\vspace{10mm}
From the lesson for the case of $n=3$, let us set up the general case. 
Let $\{\Sigma_{1},\Sigma_{3}\}$ be generalized Pauli matrices
\begin{equation}
\label{Sigma-1}
\Sigma_{1}=
\left(
\begin{array}{cccccc}
0 &   &        &        &   & 1  \\
1 & 0 &        &        &   &    \\
  & 1 & 0      &        &   &    \\
  &   & \ddots & \ddots &   &    \\
  &   &        & 1      & 0 &    \\
  &   &        &        & 1 & 0
\end{array}
\right),      \qquad 
\label{Sigma-3}
\Sigma_{3}=
\left(
\begin{array}{cccccc}
1 &        &            &        &                &               \\
  & \sigma &            &        &                &               \\
  &        & {\sigma}^2 &        &                &               \\
  &        &            & \ddots &                &               \\
  &        &            &        & {\sigma}^{n-2} &               \\
  &        &            &        &                & {\sigma}^{n-1}
\end{array}
\right)
\end{equation}
where $\sigma$ is a primitive element $\sigma=\mbox{exp}(\frac{2\pi i}{n})$ 
which satisfies 
\begin{equation}
\label{eq:simple-relations (general)}
\sigma^{n}=1,\quad \bar{\sigma}=\sigma^{n-1},\quad 
1+\sigma+\cdots +\sigma^{n-1}=0.
\end{equation}

\par \noindent
Then it is easy to see 
\begin{equation}
\Sigma_{1}^{n}=\Sigma_{3}^{n}={\bf 1}_{n},\quad 
\Sigma_{1}^{\dagger}=\Sigma_{1}^{n-1}, \quad 
\Sigma_{3}^{\dagger}=\Sigma_{3}^{n-1}, \quad 
\Sigma_{3}\Sigma_{1}=\sigma \Sigma_{1}\Sigma_{3}.
\end{equation}

If we define a Vandermonde matrix $W$ based on $\sigma$ as 
\begin{equation}
\label{eq:n-Walsh-Hadamard}
W=\frac{1}{\sqrt{n}}
\left(
\begin{array}{cccccc}
1 & 1            & 1               & \cdots & 1 & 1                    \\
1 & \sigma^{n-1} & \sigma^{2(n-1)} & \cdots & \sigma^{(n-2)(n-1)} & 
\sigma^{(n-1)^2} \\
1 & \sigma^{n-2} & \sigma^{2(n-2)} & \cdots & \sigma^{(n-2)^2} & 
\sigma^{(n-1)(n-2)} \\
\vdots & \vdots  & \vdots &        & \vdots & \vdots                   \\
1& \sigma^{2} & \sigma^{4}& \cdots & \sigma^{2(n-2)} & \sigma^{2(n-1)} \\
1& \sigma & \sigma^{2} & \cdots & \sigma^{n-2} & \sigma^{n-1}
\end{array}
\right), 
\end{equation}
then it is not difficult to see 
\begin{equation}
\label{eq:n-diagonalization}
\Sigma_{1}=W\Sigma_{3}W^{\dagger}=W\Sigma_{3}W^{-1}.
\end{equation}

\par \noindent 
That is, $\Sigma_{1}$ can be diagonalized by making use of $W$. 

\vspace{3mm}
We set
\begin{equation}
c_{j}(x)=\sum_{k=0}^{\infty}\frac{x^{kn+j}}{(kn+j)!}
\end{equation}
for $0\leq j \leq n-1$. It is of course
\[
\mbox{e}^{x}=\sum_{k=0}^{\infty}\frac{x^{k}}{k!}
=\sum_{j=0}^{n-1}c_{j}(x)
\]
and easy to see
\begin{eqnarray}
\label{eq:fundamental identity general}
\mbox{e}^{x\Sigma_{1}}
&=&
c_{0}(x){\bf 1}_{n}+c_{1}(x)\Sigma_{1}+c_{2}(x)\Sigma_{1}^{2}+\cdots 
+c_{n-2}(x)\Sigma_{1}^{n-2}+c_{n-1}(x)\Sigma_{1}^{n-1}  \nonumber \\
&=&
\left(
\begin{array}{cccccc}
c_{0}(x)   & c_{n-1}(x) &            & \cdots   & c_{2}(x) & c_{1}(x)   \\
c_{1}(x)   & c_{0}(x)   & c_{n-1}(x) & \cdots   &          & c_{2}(x)   \\
           & c_{1}(x)   & c_{0}(x)   & c_{n-1}(x) &        &            \\
\vdots     & \vdots     & \ddots     & \ddots   & \ddots   & \vdots     \\
c_{n-2}(x) &            & \cdots     & c_{1}(x) & c_{0}(x) & c_{n-1}(x) \\
c_{n-1}(x) & c_{n-2}(x) & \cdots     & c_{2}(x) & c_{1}(x) & c_{0}(x)
\end{array}
\right).
\end{eqnarray}

Let us look for the fundamental equation that $\{c_{0}(x),c_{1}(x),\cdots,
c_{n-2}(x),c_{n-1}(x)\}$ satisfy. By use of (\ref{eq:n-diagonalization}) 
\[
\mbox{e}^{x\Sigma_{1}}=\mbox{e}^{xW\Sigma_{3}W^{\dagger}}=
W\mbox{e}^{x\Sigma_{3}}W^{\dagger}
\]
we have
\begin{eqnarray}
\label{eq:fundamental identity general}
&&\left|
\begin{array}{cccccc}
c_{0}(x)   & c_{n-1}(x) &            & \cdots   & c_{2}(x) & c_{1}(x)   \\
c_{1}(x)   & c_{0}(x)   & c_{n-1}(x) & \cdots   &          & c_{2}(x)   \\
           & c_{1}(x)   & c_{0}(x)   & c_{n-1}(x) &        &            \\
\vdots     & \vdots     & \ddots     & \ddots   & \ddots   & \vdots     \\
c_{n-2}(x) &            & \cdots     & c_{1}(x) & c_{0}(x) & c_{n-1}(x) \\
c_{n-1}(x) & c_{n-2}(x) & \cdots     & c_{2}(x) & c_{1}(x) & c_{0}(x)
\end{array}
\right|      \nonumber \\
=
&&\left|
\begin{array}{cccccc}
\mbox{e}^{x} &       &           &      &      &                \\
 & \mbox{e}^{x\sigma} &           &      &      &               \\
 &       & \mbox{e}^{x\sigma^2} &      &      &                 \\
 &       &           & \ddots &      &                          \\
 &       &           &       & \mbox{e}^{x\sigma^{n-2}} &       \\
 &       &           &       &      & \mbox{e}^{x\sigma^{n-1}}
\end{array}
\right|
=\mbox{e}^{x(1+\sigma+\sigma^{2}+\cdots+\sigma^{n-2}+\sigma^{n-1})} 
=\mbox{e}^{0}=1
\end{eqnarray}
because $W$ is unitary ($|W|=1$). For example
\begin{eqnarray}
&&n=2\quad c_{0}^{2}(x)-c_{1}^{2}(x)=1\quad 
(\Longleftarrow\ (\ref{eq:fundamental identity two})) \nonumber \\
&&n=3\quad c_{0}^{3}(x)+c_{1}^{3}(x)+c_{2}^{3}(x)-3c_{0}(x)c_{1}(x)c_{2}(x)=1
\quad (\Longleftarrow\ (\ref{eq:fundamental identity three})) \nonumber \\
&&n=4\quad c_{0}^{4}(x)-c_{1}^{4}(x)+c_{2}^{4}(x)-c_{3}^{4}(x)
-2c_{0}^{2}(x)c_{2}^{2}(x)+2c_{1}^{2}(x)c_{3}^{2}(x) \nonumber \\
&&\qquad \ \ -4c_{0}^{2}(x)c_{1}(x)c_{3}(x)+4c_{0}(x)c_{1}^{2}(x)c_{2}(x)
-4c_{1}(x)c_{2}^{2}(x)c_{3}(x)+4c_{0}(x)c_{2}(x)c_{3}^{2}(x) =1. \nonumber \\
&&{}
\end{eqnarray}
We call $\{c_{0}(x),c_{1}(x),\cdots,c_{n-1}(x)\}$ the super hyperbolic 
system. 

The addition formulas are given by the simple equation
\[
\mbox{e}^{x\Sigma_{1}}\mbox{e}^{y\Sigma_{1}}
=\mbox{e}^{(x+y)\Sigma_{1}}
\]
and become
\begin{equation}
c_{j}(x+y)=\sum_{k+l=j\ (\mbox{mod}\ n)}c_{k}(x)c_{l}(y)
\quad \mbox{for}\quad 0\leq j\leq n-1.
\end{equation}
More explicitly,
\[
c_{j}(x+y)=c_{0}(x)c_{j}(y)+c_{1}(x)c_{j-1}(y)+\cdots+c_{j}(x)c_{0}(y)+
c_{j+1}(x)c_{n-1}(y)+\cdots+c_{n-1}(x)c_{j+1}(y).
\]

The new relations are given by the simple equation
\[
\mbox{e}^{x\Sigma_{1}}\mbox{e}^{y\Sigma_{1}^{\dagger}}
=\mbox{e}^{x\Sigma_{1}+y\Sigma_{1}^{\dagger}}
\]
and become
\begin{equation}
\sum_{k=0}^{j-1}c_{k}(x)c_{n-j+k}(y)+
\sum_{k=j}^{n-1}c_{k}(x)c_{k-j}(y)
=
\frac{1}{n}\sum_{k=0}^{n-1}\sigma^{k(n-j)}
\mbox{e}^{x\sigma^{k}+y\sigma^{n-k}}
\end{equation}
for $0\leq j\leq n-1$.

The generating matrix of modified Bessel functions of integer order is given 
by
\begin{equation}
\mbox{e}^{\frac{x}{2}\left(w\Sigma_{1}+\frac{1}{w}\Sigma_{1}^{\dagger}\right)}
=\sum_{k\in \seisu}I_{k}(x)w^{k}\Sigma_{1}^{k}
\end{equation}
and from this we have
\begin{equation}
\frac{1}{n}\mbox{tr}\left\{
\mbox{e}^{\frac{x}{2}\left(w\Sigma_{1}+\frac{1}{w}\Sigma_{1}^{\dagger}\right)}
\Sigma_{1}^{j}\right\}
=
\frac{1}{n}
\sum_{l=0}^{n-1}
\sigma^{lj}\mbox{e}^{\frac{x}{2}(w\sigma^{l}+\frac{1}{w}\sigma^{-l})}
=
\sum_{k\in \seisu}I_{nk-j}(x)w^{nk-j}
\end{equation}
for $0\leq j\leq n-1$. 

The result in the case of $j=0$ is known in \cite{BC} and \cite{C}.

\par \vspace{10mm}
We want to take a (formal) limit $n \longrightarrow \infty$. That is, what is 
$\Sigma_{1}\longrightarrow\ ?,\ \ \Sigma_{3}\longrightarrow\ ?$ 
It is of course impossible to take a limit with this form. For that let us 
make a small change. We set $n=2N+1$ and
\begin{equation}
\widetilde{\Sigma}_{1}=\Sigma_{1},\quad 
\widetilde{\Sigma}_{3}
=
\left(
\begin{array}{ccccccc}
\sigma^{-N} &        &           &        &        &        &      \\
  & \ddots &            &        &        &        &               \\
  &        & \sigma^{-1} &        &       &        &               \\
  &        &            & 1      &        &        &               \\
  &        &            &        & \sigma &        &               \\
  &        &            &        &        & \ddots &               \\
  &        &            &        &        &        & {\sigma}^{N}
\end{array}
\right)
\end{equation}
where $\sigma=\exp(\frac{2\pi i}{2N+1})$. Here we rewrite 
$\widetilde{\Sigma}_{3}$ as $\widetilde{\Sigma}_{3}=
\exp(\frac{2\pi i}{2N+1}\widetilde{G})$ where
\begin{equation}
\widetilde{G}=
\left(
\begin{array}{ccccccc}
-N &        &    &   &   &        &      \\
   & \ddots &    &   &   &        &      \\
   &        & -1 &   &   &        &      \\
   &        &    & 0 &   &        &      \\
   &        &    &   & 1 &        &      \\
   &        &    &   &   & \ddots &      \\
   &        &    &   &   &        & N
\end{array}
\right).
\end{equation}
The commutator $[\widetilde{G},\widetilde{\Sigma}_{1}]$ becomes
\[
[\widetilde{G},\widetilde{\Sigma}_{1}]
=
\left(
\begin{array}{cccccc}
0 &   &        &        &   & -2N  \\
1 & 0 &        &        &   &      \\
  & 1 & 0      &        &   &      \\
  &   & \ddots & \ddots &   &      \\
  &   &        & 1      & 0 &      \\
  &   &        &        & 1 & 0
\end{array}
\right)
=
\left(
\begin{array}{cccccc}
0 &   &        &        &   & 1  \\
1 & 0 &        &        &   &    \\
  & 1 & 0      &        &   &    \\
  &   & \ddots & \ddots &   &    \\
  &   &        & 1      & 0 &    \\
  &   &        &        & 1 & 0
\end{array}
\right)\quad (\mbox{mod}\ 2N+1).
\]
That is, we have the relation
\begin{equation}
[\widetilde{G},\widetilde{\Sigma}_{1}]
=\widetilde{\Sigma}_{1}\quad (\mbox{mod}\ 2N+1).
\end{equation}
In this stage, it may be better to write the (finite dimensional) Hilbert 
space as
\[
\fukuso^{2N+1}=\mbox{Vect}_{\fukuso}\{\ket{-N},\cdots,\ket{-1},\ket{0},
\ket{1},\cdots,\ket{N}\}
\]
because $\widetilde{G}\ket{n}=n\ket{n}$.

Now, if we take a formal limit $N\ \longrightarrow\ \infty$ then 
we have the fundamental relation
\begin{equation}
\label{eq:fundamental relation on QM}
[G,W]=W
\end{equation}
where
\begin{eqnarray}
G&=&
\left(
\begin{array}{ccccccc}
\ddots &    &    &   &   &   &        \\
       & -2 &    &   &   &   &        \\
       &    & -1 &   &   &   &        \\
       &    &    & 0 &   &   &        \\
       &    &    &   & 1 &   &        \\
       &    &    &   &   & 2 &        \\
       &    &    &   &   &   & \ddots 
\end{array}
\right),\quad
W=
\left(
\begin{array}{ccccccc}
\ddots &   &   &   &   &        &        \\
\ddots & 0 &   &   &   &        &        \\
       & 1 & 0 &   &   &        &        \\
       &   & 1 & 0 &   &        &        \\
       &   &   & 1 & 0 &        &        \\
       &   &   &   & 1 & 0      &        \\
       &   &   &   &   & \ddots & \ddots
\end{array}
\right),  \nonumber \\
&&{}
\end{eqnarray}
where the notations $\{G,W\}$ in \cite{OK} were used. Note that $G$ is 
a hermitian operator and $W$ a unitary operator on the Hilbert space
\begin{equation}
{\cal L}^{2}(\seisu)
=\left\{\sum_{n\in \seisu}c_{n}\ket{n}\ |\ 
\sum_{n\in \seisu}|c_{n}|^{2} < \infty \right\};\qquad 
W\ket{n}=\ket{n+1},\quad G\ket{n}=n\ket{n}.
\end{equation}
The relation (\ref{eq:fundamental relation on QM}) is just the fundamental 
one in quantum mechanice on the circle developed by Ohnuki and Kitakado 
\cite{OK}. 

A comment is in order. There is some freedom on the choice of $G$. That is, 
if we choose $G$ like
\[
G\ \longrightarrow\ G+\alpha{\bf 1},\quad 0\leq \alpha < 1
\]
the relation (\ref{eq:fundamental relation on QM}) still holds. $\alpha$ is 
interpreted as a kind of abelian gauge induced in quantum mechanice on the 
circle. 

Therefore it may be better to write the generators 
$\{G_{\alpha}\equiv G+\alpha{\bf 1},W\}$ in place of $\{G,W\}$ in \cite{OK}. 
We don't repeat the contents, so see \cite{OK} and its references.

Readers may find many interesting problems from the paper. For example, we 
can consider the {\bf generating operator}
\[
\mbox{e}^{\frac{x}{2}\left(wW+\frac{1}{w}W^{\dagger}\right)}.
\]
We leave some calculations to readers.

\vspace{5mm}
In this paper we developed the super hyperbolic structure for (all) finite 
quantum systems, and defined the generating matrix for the modified Bessel 
functions of integer order and obtained some interesting results. 
We also gave a connection to quantum mechanics on the circle by Ohnuki and 
Kitakado by taking a skillful limit. 

Our motivation is to apply the development in the paper to qudit theory 
based on finite quantum systems, which will be reported in another paper.

\vspace{5mm}
\noindent{\em Acknowledgment.}\\
The author wishes to thank K. Funahashi for helpful comments and suggestions.

%%%%%%%%%%%%%
%References%
%%%%%%%%%%%%%

\end{document}